# Finding the mass-energy counterpart to the transverse Doppler shift:

## Avoiding de Broglie's paradox

Steven M Taylor

This paper identifies specific angles of emission and reception of light for which there exists a mass-energy counterpart to the well-known transverse Doppler shift. At these angles, the relationship of proper and relative frequency is the same as that for proper and relative mass-energy of a source. Paradoxically, the transverse Doppler shift is often used to demonstrate that for specific angles of emission and reception the relationship of proper and relative frequency of light is the same as that for proper and relative time. But by carefully defining angles, this apparent paradox is resolved.

**Introduction**

When isolated, the effect of time dilation on a moving source would decrease the frequency of the light emitted when observed in an alternate inertial frame.[1] The transverse Doppler shift, which is often used to demonstrate this effect, is described as occurring when an observer receives an electromagnetic wave emitted from a source at $\theta = \pi/2$ to the relative motion.[2] The relative frequency under such conditions is $f' = f\sqrt{1-\beta^2}$; $\beta = v/c$ and with $f$ standing for frequency and $v$ as the velocity of the source.

Yet, light is also thought of as a photon.[3] Moreover; it can be argued that the energy of a photon is directly related to the amount of mass-energy that is used in creating the photon. This is discussed further in *Kinetic Energy Considerations for Orthogonal Emissions*. The frequency of a photon can be expressed as the quotient of the photon's energy and Plank's constant[4] such that $f = e/h$. As per Einstein, energy can be related to mass such that $f = mc^2/h$. Since as determined from an alternate inertial frame relative mass[5] equals $m\gamma$, with $\gamma$ defined as $\sqrt{1-\frac{v^2}{c^2}}^{-1}$, it follows that $f' = \frac{\gamma mc^2}{h}$. Noting that the inertial mass associated with the production of the photon is the product of its frequency and Plank's constant and then dividing by the square of the speed of light yields $f' = \frac{\gamma c^2}{h} fhc^{-2}$. Canceling terms produces $f' = f\gamma$. This is the relative frequency of a photon based solely on the increase in mass-energy of its source. Paradoxically, this is also the inverse of the transverse Doppler shift.

This is essentially the same paradox that de Broglie faced during his seminal work in 1923 that lead to the defining of the relationship of a particle's wavelength with the quotient of Plank's constant and its momentum, such that $\lambda = h/p$.[6] It will be shown that this apparent paradox can also be resolved for relativistic Doppler shift by clarifying the role of relativistic mass in the emission of a photon as well as carefully defining the angles of emission and reception of the light energy.

2**A Box of Light**
In his 1905 paper linking inertial mass to energy, Einstein states that "if a body gives off the energy $l$ in the form of radiation, its mass diminishes by $l/c^2$." [7] It is logical that the mass $l/c^2$ would travel at the speed of the radiation away from the center of mass of the system emitting it. Further, Einstein concluded the paper with the statement that, "If the theory agrees with the facts, then radiation carries inertia between emitting and absorbing bodies." [8]

This concept of light carrying mass combined with the law of conservation of momentum can be used to calculate the relative frequency of a photon emitted orthogonal to a sources' relativistic-velocity vector. Consider a hypothetical box with a mirrored interior. Assume that a photon is traveling at right angles to two of the walls with parallel reflecting surfaces. Through a series of reflections the box will vibrate slightly (v<<c) while conserving the position of the center of mass of the photon-box system. The photon's path is along what will be designated the x-axis (figure 1). The magnitude of the momentum of the box between collisions is the product of the box's mass and velocity of the box's center of mass. To conserve the photon-box system's center of mass position, the magnitude of the photon's momentum $hf/c$ must equal the magnitude of the box's momentum. This means that the product of the change in position of the box's center of mass and mass of the box must equal the product of the change in position of the photon and the photon's mass:

$$\Delta x_{box} m_{box} = \Delta x_{photon} \frac{hf}{c^2}  \qquad \text{Equation 1}$$

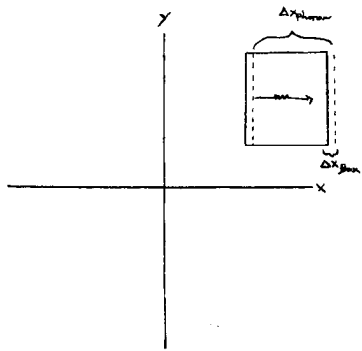

Figure 1

The conservation of momentum of the box-photon system during the reflection of the photon along the x-axis should also hold in the inertial frame of an observer traveling at a uniform velocity perpendicular to the x-axis. In such a scenario the velocity of the box could be designated as parallel to the y/y'-axis, which is also orthogonal to the x/x'-axis. (Figure 2)

According to the principle of relativity, uniform motion of inertial frames cannot be distinguished by any physical test within them.[9] Consequently the frequency of light created within an inertial frame by a process of a given energy would always produce a photon of a given frequency as measured within that inertial frame. In this way frequency can be thought of in an analogous fashion to proper time or proper rest mass. If this were not so, it would be possible to detect uniform motion of an inertial reference frame by observing the frequency of light generated within it. Therefore, the expected proper frequency of light associated with a given energy would be reflecting between the walls of the mirrored box when viewed from within the inertial frame of the box-photon system. But an observer with a uniform relativistic velocity along the y/y' axis would not calculate or observe (if emitted) the proper frequency of the photon.

To explore this, assume that the mass of the box would be sufficiently large that the velocity along the x/x' axis resulting from the reflection of the photon may be such that v<<c. Moreover, the velocity of the system relative to the observer is relativistic. As in the box-photon system's inertial frame, the overall momentum of the box-photon system in the observer's inertial frame must also be conserved during the reflection of the photon. This means that the product of the change in position and mass of the box must equal the product of the change in position and mass of the photon, yet the relativistic mass of the box will be higher by a factor of $\gamma$.[10]

$$\Delta x'_{box}\, \gamma m_{box} = \Delta x'_{photon}\, \frac{hf'}{c^2} \qquad \text{Equation 2}$$

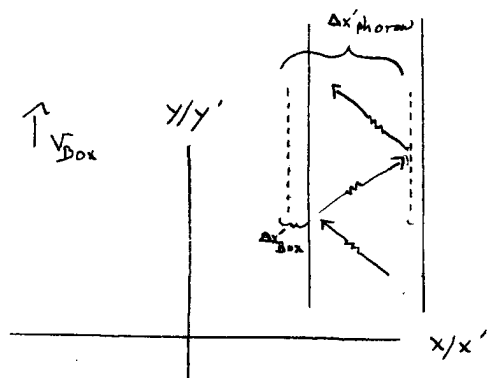

Figure 2

Since the velocity component of the photon-box system along the x/x'-axis is such that v<<c, and since length contraction does not occur along dimensions perpendicular to the line of relativistic motion, a change in position along the x-axis would equal a change in position along the x'-axis.[11] In other words, the amplitude of motion by the box and the photon along the x-axis equals the amplitude of the box and photon along the x'-axis, which means that $\Delta x' = \Delta x$. Taking the quotient of equations 2 and 3 yields:

$$f' = f\gamma. \qquad \text{Equation 3}$$

Thus, in the observer's inertial frame a photon of frequency $f\gamma$ is reflecting between the walls of a box of mass $m\gamma$. This result is consistent with equation one, where the relative mass energy of a source is used to calculate the relative frequency of a photon.

## Mass-Energy Counterpart to the Transverse Doppler Shift

The box-photon model demonstrates that for an observer to intercept a photon of relative frequency $f\gamma$, it must have been emitted from a source at an angle orthogonal to the y/y'-axis as seen within the system's inertial frame. Such orthogonal emissions relative to the source's inertial frame correspond to $\theta = \pi/2$ and $3\pi/2$ radians as used in the all-angle relativistic Doppler shift equation: $f' = f\gamma(1 - \beta\cos\theta)$.[12] As will be discussed later, it is important to note that Einstein defines θ as formed by "the connecting line 'light source-observer' with the *observer's velocity*."[13] These angles correspond to the mass-energy counterpart to the traditional transverse Doppler shift where the relative frequency of a photon can be attributed solely to the increase in the relative mass-energy of a source.

## Kinetic Energy Considerations for Orthogonal Emissions

As demonstrated in the photon-box model, the relative frequency of a photon emitted orthogonally to a sources velocity (relative to the source's inertial frame) is given by $f' = \gamma mc^2 / h$, where m is the amount of proper mass-energy of the photon. Light emitted orthogonal to the source's velocity (relative to the sources' inertial frame) will be intercepted by an observer with a higher frequency. That difference in frequency is $(f\gamma - f)$. Therefore the difference in relative and proper frequency for an orthogonal emission is $f\gamma - f = \dfrac{\gamma mc^2}{h} - \dfrac{mc^2}{h}$. Solving for the difference in energy of the relative and proper frequency yields $(f\gamma - f)h = \gamma mc^2 - mc^2$. Kinetic energy is also defined as the difference in total energy and the product of proper-mass and the square of the speed of light $Ke = \gamma mc^2 - mc^2$.[14] Setting these equations equal yields $Ke = (f\gamma - f)h$. This means that the difference in energy of the relative and proper frequency of an orthogonally emitted photon equals the kinetic energy of the proper mass-energy that created the photon prior to its emission.

Credence is lent to this interpretation by noting that in his 1905 paper linking the inertia of a body to its energy, Einstein noted that the difference in the kinetic energy of a moving source can be equated to the product of the energy of the emission and the square of the source velocity, divided by two times the speed of light squared.[15]

Using Einstein's notation $k_0 - k_1 = \dfrac{l}{V^2}\dfrac{v^2}{2}$ with $V$ representing the speed of light.

For the specific case of an orthogonal emission, the kinetic energy of the photon's proper mass-energy equals precisely the increase in energy of the relative frequency of the photon with respect to the energy of its proper frequency. Since energy is conserved, the post-emission system will have the same ratio of relative mass-energy to proper mass-energy $\dfrac{\gamma' m'c^2}{m'c^2} = \dfrac{\gamma mc^2}{mc^2}$ with $m'$ being the proper mass of the system after emission;

therefore $\gamma' = \gamma$. Consequently the center of mass of the box will have the same magnitude of translational velocity as the box-photon system prior to emission.

**Longitudinal Doppler shift**
Relativistic Doppler shift equations for longitudinal emissions are often derived using an electromagnetic wave interpretation for light.[16] Yet, an analysis using conservations of momentum and energy produces a derivation for longitudinal emissions.[17]

Consider a system of mass $M$ traveling at velocity $v$ away from an observer at $t_0$. At $t_1$ photon is emitted toward the observer ($\theta = 0$).

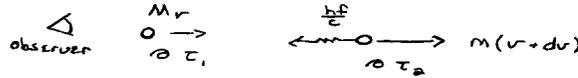

Figure 3

Due to the conservation of momentum, the sum of the photon's momentum equals the momentum of the system prior to emission such that $-hf/c = (M-m)v - mdv$.
Although not exact $M - m \approx (M_0 - m_0)\gamma$ which yields $-hf/c = (M_0 - m_0)\gamma v - mdv$.

Moreover, $(M_0 - m_0)\gamma v - mdv = \dfrac{hf_0}{c^2}\gamma v - \gamma m_0 dv$ yielding $-\dfrac{hf}{c} = \dfrac{hf_0}{c^2}\gamma v - \gamma m_0 dv$.

Since $m_0 dv = \dfrac{hf_0}{c} \rightarrow -\dfrac{hf}{c}\sqrt{1-\beta^2} = \dfrac{hf_0}{c^2}v - \dfrac{hf_0}{c}$ which yields for a receding

source $f = f_0(1-\beta/1+\beta)^{1/2}$ (equation A) and for an approaching source

$f = f_0(1+\beta/1-\beta)^{1/2}$ (equation B). This is what the general relativistic Doppler shift equation reduces to for $\theta = 0$ and $\theta = \pi$ respectively. Thus, using the photon-box model, a photon would reflect back and forth along the y/y' -axis and could be thought of as alternating between the frequencies given by equations A and B as seen by an observer in an alternate inertial frame.

**Kinetic Energy Considerations for Non-Orthogonal Emissions**
As demonstrated, $(f\gamma - f)$ equals the kinetic energy of a photons mass energy (relative to the observer) prior to its emission. Since $f\gamma$ represents the total energy within the box-photon system associated with the mass-energy that created the photon, any value for a relative frequency other than $f\gamma$ reflects a change in the ratio of total energy and kinetic energy of the box.

This means that a non-orthogonal emission (relative to the source's inertial frame) of a photon causes a change in magnitude of the translational kinetic energy of the system emitting the photon. The precise value of which is $\Delta Ke = (f'-f\gamma)h$ where $f'$ is the observed relative frequency. When the value of $(f'-f\gamma)$ is positive, the increased energy and frequency of the photon represents a reduction in the ratio of the relative mass-energy to proper mass-energy of the box system after emission: $\dfrac{\gamma' m' c^2}{m' c^2} < \dfrac{\gamma m c^2}{m c^2}$, therefore $\gamma' < \gamma$. This represents a lower translational velocity for the box's center of mass relative to the observer's inertial frame. For negative values an emission would result in a higher translational kinetic energy for the box.

**Avoiding Paradox by Using the Correct Theta**
By carefully defining and keeping track of angles of emission and reception it is possible to eliminate any paradox regarding the effects of relative motion upon the frequency of light emitted from a moving source and or observer.

As previously noted, Einstein defined the angle of emission as the angle θ that the line 'light-source observer' forms with the *observer's velocity*. As demonstrated with the photon-box system, emissions at θ = π/2 and θ = 3π/2 provide the mass-energy counterpart to the traditional transverse Doppler shift. It is with such orthogonal emissions the relationship of proper and relative frequency stand in the exact relationship as proper and relative mass-energy, such that $f' = f\gamma$.

As noted, a common interpretation of the transverse Doppler shift uses the electromagnetic wave interpretation to demonstrate the purely relativistic effect of time upon light's frequency. The relationship for relative frequency associated with the transverse Doppler shift, $f' = f\sqrt{1-\beta^2}$ also involves right angles, but the right angle is formed by the line 'observer-light source' with the *source's velocity*.[18]

For the purpose of clarity, angles formed by the line "observer-light source" with a source's velocity will be denoted as θ ', and those that are formed by the line "observer-light source" with the observer's velocity will be denoted as θ.

**Putting It All Together**
Consider two mirrors oriented so that a light signal may be reflected between their parallel surface planes, while the mirrors are moving at a relativistic velocity toward each other. As the mirrors move closer one mirror emits (or reflects) a light signal at an angle of θ = π/2, which impinges on the surface of the second mirror.

Designating the mirror which originates the signal as A, and the other B, it is clear that the sequence of events can be viewed with equal validity from both A's inertial frame and B's inertial frame.

As seen from A's inertial frame the transmission can be interpreted as an electromagnetic transmission to an observer. As such, the relativistic effect of time dilation alone can be used to predict that within B's inertial frame a relative frequency of $f' = f\gamma$.

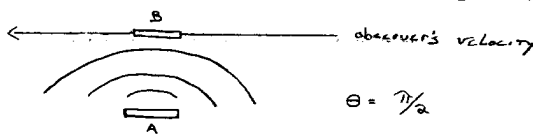

Figure 4

$f' = \dfrac{1}{t'}$ since $t' = t\gamma^{-1} \rightarrow f' = f\dfrac{1}{\gamma^{-1}}$ which yields $f' = f\gamma$.

In this scenario, the ratio of relative frequency and proper frequency is the same as that as relative and proper time. This can be interpreted as the source clock running at a faster rate than the observer's clock and using an electromagnetic wave interpretation for the light signal.

In B's inertial frame, the same transmission has an angle formed by the line light-source-observer with he source's velocity of $\theta' = \gamma \arctan \beta^{-1}$, where $\beta = \frac{v}{c}$.[19]

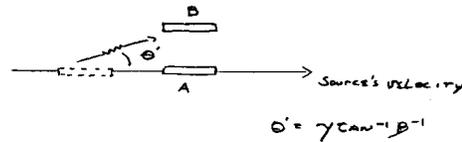

Figure 5

From this perspective, the increase in frequency of the light signal can be attributed solely to the higher total energy to rest-mass ratio of the source.

$f' = e'/h$, and since $e' = \gamma m c^2 \rightarrow f' = \frac{\gamma m c^2}{h}$. Given that $m = \frac{hf}{c^2} \rightarrow f' = \frac{hf}{c^2} \gamma c^2 h^{-1}$ yielding $f' = f\gamma$.

At this angle, the ratio of relative and proper frequency is the same as that of relative and proper mass-energy, and represents the mass-energy counterpart to the transverse Doppler shift. Conveniently, $f' = f\gamma$ in both A's and B's inertial frames.

From A's inertial frame, the return signal can be viewed as B returning a frequency of $f\gamma$ relative to B's inertial frame. Yet, correcting for B's time dilation the electromagnetic wave will be measured as the frequency of light originally sent to B.

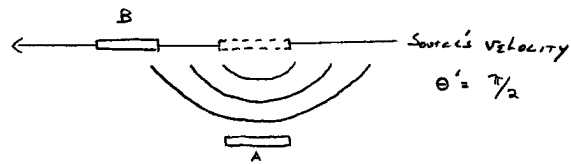

Figure 6

$f'' = \frac{1}{t''}$ since $t'' = t'\gamma \rightarrow f'' = f' \frac{1}{\gamma}$ giving $f'' = f'\gamma^{-1}$ and given

that $f' = f\gamma \rightarrow f'' = f\gamma\gamma^{-1} = f$.

This is essentially the transverse Doppler shift with the source having the slower clock while it emits the light wave.

Since the angle of incidence equals the angle of reflection, from the perspective of B's inertial frame, the return signal will be reflected back to A at the same angle that it was received. When viewed as a photon from B's inertial frame, the transmission is to a system of a higher total energy to rest mass-energy ratio. Relative to A's inertial frame, the impinging photon will be orthogonal to A's velocity. Symmetry suggests that according to A's inertial frame the photon will be of the same frequency that was sent.

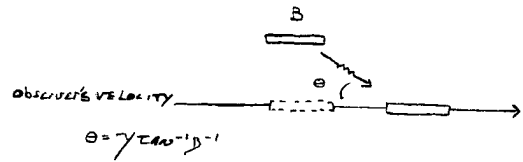

Figure 7

$f'' = \dfrac{e''}{h}$ since $e'' = \gamma m' c^2 \rightarrow f'' = \dfrac{\gamma m' c^2}{h}$. And given that $m' = m\gamma^{-1} \rightarrow f'' = \dfrac{\gamma m \gamma^{-1} c^2}{h}$ yielding $f'' = \dfrac{mc^2}{h} = f$.

## Conclusion

By carefully defining the angles of emission and reception of light energy from a moving source it is easy to see how the relativistic effects of motion affect both the wave and mass-energy aspects of light without paradox.

Emissions orthogonal to the source line of motion (relative to the source's inertial frame) and of the angle, $\theta' = \gamma \arctan \beta^{-1}$ (as seen from the observer's inertial frame) are the mass-energy counterpart to the transverse Doppler shift that is often used to demonstrate the isolated effects of time dilation on a transverse wave. With orthogonal emissions, proper and relative frequency of a photon stands in the same relationship as proper and relative mass-energy.

With the mass-energy counterpart to the transverse Doppler shift, the increase in energy of the photon has precisely the value of kinetic energy of the mass-energy which created the photon prior to emission from a system. This results in a post-emission system (that does not include the emitted photon) that has the same magnitude of relative velocity and thus the same ratio of relative to proper mass. Any other angle of emission either increases or decreases the ratio of relative to proper mass, and in essence increases or lessens the magnitude of the system's subsequent kinetic energy and magnitude of velocity as well.

Given the conceptual importance of the dual nature of light and the special theory of relativity as well as the Plank-Einstein formula equating frequency and energy of a photon, the identification and analysis of the mass-energy counterpart to the well-known transverse Doppler shift could help students better understand the complementary nature of basic quantum theory and the Special Theory of Relativity.

## Endnotes

[1] R. Resnick, *Relativity and Early Quantum Theory*, (John Wiley and Sons inc.1972) pg. 68
[2] Ibid